\documentclass[prd,aps,showpacs,nofootinbib,tightenlines]{revtex4}
\usepackage{amsmath}
\usepackage{bm}
\usepackage{times}
\usepackage{braket}
\usepackage{color}
\usepackage{epsfig}
\usepackage{slashed}
\usepackage{hyperref}
\usepackage{amssymb}
\usepackage{graphicx}
\newcommand{\beq}{\begin{eqnarray}}
\newcommand{\eeq}{\end{eqnarray}}
\newcommand{\non}{\nonumber\\ }

\def \cpc{ Chin. Phys. C  }

\def \epjc{ Eur. Phys. J. C }
\def \ijmpa{ Int. J. Mod. Phys. A }
\def \jpg{  J. Phys. G }
\def \npb{  Nucl. Phys. B }
\def \plb{  Phys. Lett. B }
\def \ppnp{ Prog.Part. $\&$ Nucl. Phys. }

\def \prd{  Phys. Rev. D }
\def \prl{  Phys. Rev. Lett.  }

\def \zpc{  Z. Phys. C }
\def \jhep{ JHEP }

\definecolor{Red}{rgb}{1.,0.,0.}

\definecolor{Blue}{rgb}{0.,0.,1.}

\definecolor{nicered}{rgb}{0.7,0.1,0.1}
\definecolor{nicegreen}{rgb}{0.1,0.5,0.1}
\hypersetup{colorlinks,citecolor=nicegreen,linkcolor=nicered}
\begin{document}

\title{Quasi-two-body decays $B \to  D  K^*(892) \to D  K \pi$ in the perturbative QCD approach}
\author{Ai-Jun Ma$^1$}  \email{theoma@163.com}
\author{Wen-Fei Wang$^2$}\email{wfwang@sxu.edu.cn}
\author{Ya Li$^3$}\email{liyakelly@163.com}
\author{Zhen-Jun Xiao$^4$}\email{xiaozhenjun@njnu.edu.cn}
%
\affiliation{$^1$ Department of Mathematics and Physics,  Nanjing Institute of Technology, Nanjing, Jiangsu 211167, P.R. China}
\affiliation{$^2$ Institute of Theoretical Physics, Shanxi University, Taiyuan, Shanxi 030006, P.R. China}
\affiliation{$^3$ Department of Physics, College of Science, Nanjing Agricultural University, Nanjing, Jiangsu 210095, P.R. China}
\affiliation{$^4$ Department of Physics and Institute of Theoretical Physics, Nanjing Normal University, Nanjing, Jiangsu 210023, P.R. China}
\date{\today}
\begin{abstract}
We study the quasi-two-body decays $B\to  D K^*(892) \to D K\pi$ by employing the perturbative QCD approach.
The two-meson distribution amplitudes $\Phi_{K\pi}^{\text{$P$-wave}}$ are adopted to describe the final state interactions
of the kaon-pion pair  in the resonance region.  The resonance line shape for the $P$-wave $K\pi$ component $K^*(892)$
in the time-like form factor $F_{K\pi}(s)$ is parameterized by the relativistic Breit-Wigner function.
For most considered decay modes,  the theoretical predictions for their branching ratios  are consistent with currently
available experimental measurements within errors.
We also disscuss  some ratios of  the branching fractions of the concerned decay
processes. More precise data from LHCb and Belle-II are expected to test our predictions.
\end{abstract}

\maketitle

\section{Introduction}\label{sec:1}
Many three-body hadronic $B$ meson decays including $B \to DK\pi$ have been studied experimentally in recent
years~\cite{PDG2018,epjc77-895}.
The $B \to DK\pi$ decays have demonstrated the potential to determine the CKM angles precisely.
Suggestions for the determination of the unitarity triangle angle $\gamma$ through Dalitz plot~\cite{pm44-1068} analyses of the
decays $B^\pm\to D K^\pm\pi^0$ and $B^0 \to DK^+\pi^-$ were proposed in~\cite{prd67-096002} and~\cite{prd79-051301,
prd80-092002}, respectively. And the measurement has been performed by LHCb~\cite{prd92-012012}.  The
${\it BABAR}$ Collaboration has presented a measurement of the weak phase $2\beta+\gamma$ from the time-dependent
Dalitz plot analysis of the $B^0 \to D^{\mp} K^0 \pi^{\pm}$ decays~\cite{prd77-071102}.
In addition, the decay modes  $B\to DK\pi$ have provided rich opportunities to investigate the spectroscopy of
excited charm mesons and the significant components originated from $K\pi$ system, corresponding results have been
acquired from Dalitz plot analyses of $B^0 \to D^{(*)\pm}K^0\pi^{\mp}$~\cite{prl95-171802},
$B_s^0 \to \bar{D}^0K^-\pi^+$~\cite{prl113-162001,prd90-072003}, $B^- \to D^+K^-\pi^-$~\cite{prd91-092002} and $B^+ \to
D^+K^+\pi^-$~\cite{prd93-051101} decays.  In the amplitude analyses of $B \to  D  K \pi$ decays, contributions from the $P$-wave
$K\pi$ resonant state $K^*(892)$\footnote{In the following sections, $K^*$ represents $K^*(892)$ without specific reference.} were
found to be the largest proportion in most cases, and the $B \to DK^*$ decays have been substantially studied in experiment by
quasi-two-body approach~\cite{prd90-112002,prd73-111104,prl90-141802,prl96-011803,prd74-031101,prd78-032005,
plb706-32,plb727-403,prd82-092006,jhep1302-043}.

On the theoretical side, the charmed hadronic $B$ meson decays $B\to DK^*$ have been studied by using  rather different methods
in Refs.~\cite{jhep1609-112,zpc34-103,plb318-549,ijmpa24-5845,prd75-074021,prd92-094016,jpg37-015002,hep27-1062,
prd78-014018}. The $K^*$ was treated as a stable particle in the framework of two-body decays while as a resonance
with the cascade decay $K^*\to K\pi$ in the three-body decays. Several approaches have been adopted to describe those
three-body $B$ decays involving $K\pi$ systems.
For instance, within the QCD factorization~\cite{prl83-1914,npb591-313,npb675-333}, the authors studied the $CP$
violation and the contribution of the strong kaon-pion interactions in the three-body $B\to K\pi\pi$ decays~\cite{prd79-094005,
prd81-094033} where the $K^*(892)$ and $K_0^*(1430)$ resonance effects were mainly taken into account.
In Ref.~\cite{1811.02167}, the calculation of the localized $CP$ violation in $B^-\to K^-\pi^+\pi^-$
decays has been done with the $K\pi$ channels including $K^*(892)$, $K_0^*(1430)$, $K^*(1410)$, $K^*(1680)$ and
$K_2^*(1430)$. Using a simple model on the basis of the factorization approach, the branching ratios and direct $CP$
violation for the charmless three-body hadronic decays $B_{(s)} \to KK\pi$ and $B_{(s)} \to K\pi\pi$  have been calculated in
Refs.~\cite{prd88-114014,prd89-074025,prd94-094015,prd89-094007}. In the recent works, the $K\pi$ contributions to the
decay channels $B\to \psi K\pi$~\cite{prd97-033006} and $B\to P K\pi$~\cite{1809.09816}, with
$\psi=(J/\psi,\psi(2S))$ and  $P=(K,\pi)$, were analyzed by employing the perturbative QCD (PQCD) factorization
approach~\cite{plb504-6,prd63-054008,prd63-074009,li2003}. In addition, phenomenological studies of the processes
$\bar{B}^0 \to D^0\pi^+\pi^-$ and $\bar{B}_s^0 \to D^0 K^+\pi^-$  based on results of the chiral unitary approach has been performed
in Ref.~\cite{prd92-034008}.  Motivated by the abundant experimental data and theoretical studies, we shall analyse the
contributions of the resonance $K^*$ in the $B \to DK\pi$ decays in this work.

In the framework of the PQCD factorization approach, the two-body charmed
hadronic decays $B \to D M$ have been studied for many years~\cite{hep27-1062,prd78-014018,jpg37-015002,
prd52-3958,prd53-4982,prd67-054028,prd69-094018,epjc24-121,epjc28-305,prd68-097502,jpg29-2115,prd86-094001,cpc37-013103,prd87-074030,
prd95-016011,epjc77-870}. In Refs.~\cite{prd52-3958,prd53-4982}, the authors examined the PQCD formalism
to $B$ to $D$ transitions and  discussed some related two-body nonleptonic decays. Assuming the hierachy $m_B\gg m_D\gg\bar{\Lambda}$
 with $\bar{\Lambda}=m_B-m_b$, the $B \to D^{(*)}$ form factors in the
 heavy-quark and large-recoil limits were calculated in Ref.~\cite{prd67-054028}. The next-to-leading-power corrections
 were found to be less than $20\%$ of the leading contribution, indicating that the power expansion made sense, and the results
of the $B\to D \pi$ branching ratios were consistent with the experimental results.
 It is worth to mention that
 the contribution from nonfactorizable and annihilation-type diagrams is also important in the $B \to DM$ decays~\cite{prd69-094018}.
 As a feature of PQCD, all topologies of decay amplitudes are calculable which makes it advantageous to study those
 charmed $B$ decays including the pure annihilation type decays and the color suppressed decays in the PQCD approach.
 Some separate calculations for the two-body charmed decays of $B$ meson were carried out in
Refs.~\cite{hep27-1062,epjc24-121,epjc28-305,prd68-097502,jpg29-2115}.
 In Ref.~\cite{epjc28-305}, specifically, the authors studied the annihilation type
 decay $B \to D_s K$ and gave the PQCD prediction for a sizable branching ratio ${\cal B}(B^0 \to D^-_s K^+)\sim10^{-5}$,
 which has already been confirmed by experiments~\cite{PDG2018,epjc77-895}. In the past decade, the two-body charmed decays
 $B\to DS, DP, DV, DA, DT$, where $S, P, V, A, T$ denote the scalar, pseudoscalar,
vector, axial-vector and tensor mesons, have
been studied systematically in~\cite{prd78-014018,jpg37-015002,prd86-094001,cpc37-013103,prd87-074030,
prd95-016011,epjc77-870} by employing the PQCD approach and most of the predictions are in good agreement with the available
experimental data. Therefore, it is interesting and meaningful to analyse the relevant three-body charmed hadronic $B$ meson
decays within the same method.

Theoretically, the three-body hadronic $B$ meson decays are much more complicated than the two-body cases because
of their non-trivial kinematics and the different phase space distributions.
While these three-body decay processes are known to be dominated by the low energy scalar, vector and tensor resonant
states, which could be handled in the quasi-two-body framework by neglecting the three-body and rescattering
effects~\cite{npb899-247,jhep1710-117,1512-09284,plb763-29,1812.08524}.  In the quasi-two-body region of the phase space, the three final states
are quasi-aligned in the rest frame of the $B$ meson and two of them almost collimate to each other,
the related processes can be denoted as $B\to h_1R\to h_1h_2h_3$ where $h_1$ represents the bachelor particle
and the $h_2h_3$ pair proceeds by the intermediate state $R$.
In our previous works, the $S$-, $P$- and $D$-wave $\pi\pi$ and $K\pi$ resonance contributions to a series of charmed or
charmless three-body $B_{(s)}$ meson decays have been studied in~\cite{prd97-033006,1809.09816,prd91-094024,
plb763-29,epjc76-675,cpc41-083105,epjc77-199,npb923-54,prd95-056008,prd96-036014,prd96-093011,npb924-745,
prd98-113003} within the PQCD approach by introducing two-meson distribution
amplitudes~\cite{Muller,Grozin,prl81-1782,npb555-231,plb561-258,prd70-054006,prd89-074031}.
The consistency of the theoretical studies and experimental results indicates the PQCD
factorization approach are applicable to the three-body and quasi-two-body hadronic $B$ meson decays.
More recently, several quasi-two-body decays involving $D^*_0(2400)$~\cite{plb788-468}, $D^*(2007)^0$ and
$D^*(2010)^\pm$~\cite{1812.08524} as the intermediate states have been studied. And the isovector scalar resonances $a_0(980)$
and $a_0(1450)$ in the $B\to \psi (K\bar K,\pi\eta)$ decays were presented in~\cite{1811.12738}.
In this work, we will extend the previous studies to the quasi-two-body decays $B\to DK^*\to DK\pi$.

This paper is organized as follows.  In Sec.~II, we give a brief introduction for the theoretical
framework and perturbative calculations for the considered decays. Then, the numerical values
and phenomenological analyses are given in Sec.~III. Finally, the last section contains a short summary.

\section{The theoretical framework}\label{sec:2}
In the framework of the PQCD approach for the quasi-two-body decays, the nonperturbative dynamics associated with the
pair of the mesons are absorbed into two-meson distribution amplitudes, then the relevant decay amplitude $\cal A$ for the
quasi-two-body decays $B \to  D  K^* \to D  K \pi$ can be written as the convolution~\cite{plb561-258,prd70-054006}
\beq
{\cal A}=\Phi_B\otimes H\otimes \Phi_{D}\otimes\Phi_{K\pi}^{\text{$P$-wave}},
\eeq
where the symbol $\otimes$ means the convolution integrations over the parton momenta and the hard kernel $H$ includes
the leading-order contributions. The $B$ meson ($D$ meson, $P$-wave $K\pi$ pair) distribution amplitude $\Phi_B$
($\Phi_{D}$,$\Phi_{K\pi}^{\text{$P$-wave}}$) absorbs the nonperturbative dynamics in the decay processes.

\subsection{Coordinates and wave functions}\label{sec:21}
In the rest frame of the $B$ meson, we define the $B$ meson momentum $p_{B}$, the kaon momentum $p_1$,
the pion momentum $p_2$,  the $K^*$ meson momentum $p=p_1+p_2$
and the $D$ meson momentum $p_3$ in the light-cone coordinates as
\beq\label{lc}
p_{B}&=&\frac{m_{B}}{\sqrt2}(1,1,\textbf{0}_{\rm T}),
~\quad p=\frac{m_{B}}{\sqrt2}(1-r^2,\eta,\textbf{0}_{\rm T}),~\quad
p_3=\frac{m_{B}}{\sqrt2}(r^2,1-\eta,\textbf{0}_{\rm T}), \non
p_1&=& \left ( \frac{m_{B}}{\sqrt2}\zeta (1-r^2), \frac{m_{B}}{\sqrt2}(1-\zeta)\eta, \textbf{p}_{1\rm T}\right), \quad
p_2=  \left( \frac{m_{B}}{\sqrt2}(1-\zeta)(1-r^2),\frac{m_{B}}{\sqrt2} \zeta\eta, \textbf{p}_{2\rm T}\right),
\eeq
with the mass ratio $r = m_{D}/m_{B}$, $m_{B(D)}$ is the mass of the $B(D)$ meson. The variable $\eta$ is defined as
$\eta=\omega^2/[(1-r^2)m^2_{B}]$ with the invariant mass squared $\omega^2=p^2=m^2(K\pi)$ of the kaon-pion pair
and $\zeta$ is the momentum fraction for the kaon meson. The momenta of the light quarks in the $B$ meson, the $K^*$
meson and the $D$ meson are chosen as $k_B$, $k$ and $k_3$ respectively
\beq
k_B=(0,x_Bp_B^-,\textbf{k}_{B \rm T}),~\quad k=(zp^+,0,\textbf{k}_{\rm T}),~\quad
k_3=(0,x_3p_3^-,\textbf{k}_{3\rm T}),
\eeq
where the  corresponding momentum fractions $x_{B}$, $z$ and $x_3$ run between zero and unity.

The $P$-wave kaon-pion distribution amplitudes are defined in the same way as in Ref.~\cite{plb561-258,prd70-054006},
\beq
\Phi_{K\pi}^{\text{$P$-wave}}=\frac{1}{\sqrt{2N_c}}[{ p \hspace{-2.0truemm}/
}\phi_0(z,\zeta,\omega^2)+\omega\phi_s(z,\zeta,\omega^2)
+\frac{{p\hspace{-1.5truemm}/}_1{p\hspace{-1.5truemm}/}_2
  -{p\hspace{-1.5truemm}/}_2{p\hspace{-1.5truemm}/}_1}{\omega(2\zeta-1)}\phi_t(z,\zeta,\omega^2)] \;,
\label{eq:phifunc}
\eeq
with the functions~\cite{1809.09816}
\beq
\phi_0&=&\frac{3F_{K\pi}(s)}{\sqrt{2N_c}} z
(1-z)\left[1+a_{1K^*}^{||}3(2z-1)+
a_{2K^*}^{||}\frac{3}{2}(5(2z-1)^2-1)\right]P_1(2\zeta-1)\;,\label{eq:phi1}\\
\phi_s&=&\frac{3F_s(s)}{2\sqrt{2N_c}}(1-2z)P_1(2\zeta-1) \;, \label{eq:phi2}\\
\phi_t&=&\frac{3F_t(s)}{2\sqrt{2N_c}}(2z-1)^2P_1(2\zeta-1) \;,
\label{eq:phi3}
\eeq
where the Legendre polynomial $P_1(2\zeta-1)=2\zeta-1$ and the variable $s=\omega^2=m^2(K\pi)$.
For the Gegenbauer moments, we adopt $a_{1K^*}^{||}=0.05\pm0.02$ and $a_{2K^*}^{||}=0.15\pm0.05$ determined
in Ref.~\cite{1809.09816}. The relativistic Breit-Wigner (RBW) function is an appropriate model for narrow resonances
which are well separated from any other resonant or nonresonant contributions with the same spin,
and it is widely used in the experimental data analyses. Here, the time-like form factor
$F_{K\pi}(s)$ is parameterized with the RBW line shape and can be expressed as the following form~\cite{prd92-012012,prd91-092002,prd90-072003}
\beq
F_{K\pi}(s)&=&\frac{m_{K^*}^2}{(m^2_{K^*} -s)-im_{K^*}\Gamma(s)},
\eeq
with the mass-dependent decay width $\Gamma(s)$
\beq
\Gamma(s)&=&\Gamma_{K^*}\frac{m_{K^*}}{\sqrt{s}}\left(\frac{|\overrightarrow{p_1}|}{|\overrightarrow{p_0}|}\right)^3
\frac{1+(|\overrightarrow{p_0}|r_{BW})^2}{1+(|\overrightarrow{p_1}|r_{BW})^2}.
\eeq
The $|\overrightarrow{p_1}|$ is the
momentum of one of the resonance daughters evaluated in the $K\pi$ rest frame and $|\overrightarrow{p_0}|$ is the value of
$|\overrightarrow{p_1}|$ when $\sqrt{s}=m_{K^*}$.
The pole mass $m_{K^*}$ and width $\Gamma_{K^*}$ of the resonance state $K^*$ are chosen as
$m_{K^{*0}}=895.55\pm0.20~(m_{K^{*\pm}}=891.76\pm0.25)$ {\rm MeV}  and $\Gamma_{K^{*0}}=47.3\pm0.5~(\Gamma_{K^{*\pm}}
=50.3\pm0.8)$ {\rm MeV}, respectively~\cite{PDG2018}. The parameter $r_{BW}$ is the barrier radius which is set to $4.0$ {\rm GeV}
$^{-1}$ as in Ref.~\cite{prd92-012012,prd91-092002,prd90-072003}. Following Ref.~\cite{plb763-29}, we also assume that
$F_s(s) = F_t(s) \approx (f_{K^*}^T/f_{K^*}) F_{K\pi}(s)$ with $f_{K^*}=0.217 \pm 0.005$ {\rm GeV} and
$f^T_{K^*}=0.185 \pm 0.010$ {\rm GeV}~\cite{prd76-074018}.

In this work, we use the same distribution amplitudes for the $B$ and $D$ meson as in Ref.~\cite{npb923-54,prd96-093011}
where one can easily find their expressions and the relevant parameters.

\subsection{Analytic formulae}\label{sec:22}
For the quasi-two-body decays $B \to  D  K^* \to D  K \pi$, the effective Hamiltonian
is defined as~\cite{rmp68-1125}
\beq
{\cal  H}_{eff}&=& \left\{\begin{array}{ll}
\frac{G_F}{\sqrt{2}}V^*_{cb}V_{ud(s)}\left[C_1(\mu)O_1(\mu)+C_2(\mu)O_2(\mu)\right],
& \ \  {\rm for} \ \ B_{(s)} \to   \bar{D}_{(s)} K^* \to \bar{D}_{(s)} K \pi\ \ {\rm decays},\\
\frac{G_F}{\sqrt{2}} V^*_{ub}V_{cd(s)}\left[C_1(\mu)O'_1(\mu)+C_2(\mu)O'_2(\mu)\right],
& \ \  {\rm for} \ \ B_{(s)} \to  D_{(s)} K^* \to D_{(s)}  K \pi\ \ {\rm decays},\\
\end{array} \right.
\eeq
where the Fermi coupling constant $G_F=1.16638\times 10^{-5}$ {\rm GeV}$^{-2}$,
$V_{ij}$ are the CKM matrix elements
and $C_{1,2}(\mu)$ denote the Wilson coefficients at the renormalization scale
$\mu$. The $O^{(\prime)}_{1,2}$ represent the effective four quark operators and can be expressed as
 \beq
O_1=(\bar{b}_\alpha   c_\beta)_{V-A} (\bar{u}_\beta   d(s)_\alpha)_{V-A},~~~~
O_2=(\bar{b}_\alpha   c_\alpha)_{V-A} (\bar{u}_\beta   d(s)_\beta)_{V-A},\label{eq:oo1}\\
O'_1=(\bar{b}_\alpha   u_\beta)_{V-A} (\bar{c}_\beta   d(s)_\alpha)_{V-A},~~~~
O'_2=(\bar{b}_\alpha   u_\alpha)_{V-A} (\bar{c}_\beta   d(s)_\beta)_{V-A},\label{eq:oo2}
\eeq
with the color indices $\alpha$ and $\beta$. Here $V-A$ refers to the Lorentz structure $\gamma_\mu(1-\gamma_5)$
and $(\bar{q}_1q_2)_{V-A}=\bar{q}_1 \gamma_\mu(1-\gamma_5)q_2$.

\begin{figure}[tbp]
\begin{center}
\vspace{-2cm}
\centerline{\epsfxsize=17cm \epsffile{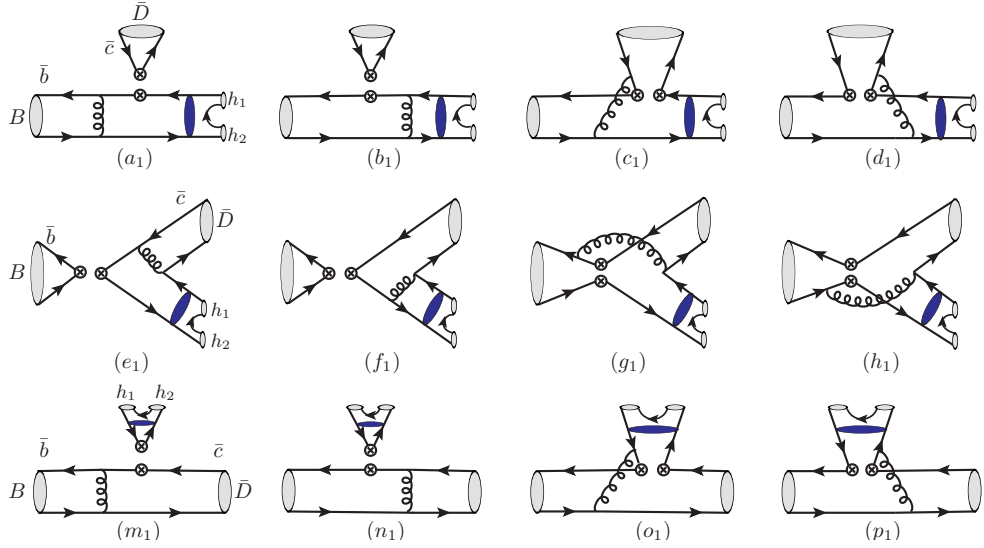}}
\vspace{-14.7cm}
\caption{The leading-order Feynman diagrams for the quasi-two-body decays $B_{(s)} \to \bar{D}_{(s)} K^* \to\bar{D}_{(s)} K\pi$
where $B_{(s)}$ = ($B^0$, $B^+$, $B^0_s$) and $\bar{D}_{(s)} $ =( $\bar{D}^0$, $D^-$, $D_s^-$). The $h_1h_2$ denotes the
$K\pi$ pair and the blue ellipse represents the intermediate state $K^*$. The ``$\otimes \otimes$'' in the diagrams represents the insertion of the 4-fermion operator $O_i$ in the effective theory~\cite{rmp68-1125}. When considering the color structure of
the diagrams for the single-gluon exchange between
two color-singlet weak-currents, the operators
$O_1$ and $O_2$, with the same flavor form but different color structure, have to be introduced. The expressions for the relevant operators
$O_{1,2}$ and $O^{\prime}_{1,2}$ (for FIG.~\ref{fig:fig2}) can be found in  Eq.~(\ref{eq:oo1}) and Eq.~(\ref{eq:oo2}).}
\label{fig:fig1}
\end{center}
\end{figure}

\begin{figure}[tbp]
\begin{center}
\vspace{-2.5cm}
\centerline{\epsfxsize=17cm \epsffile{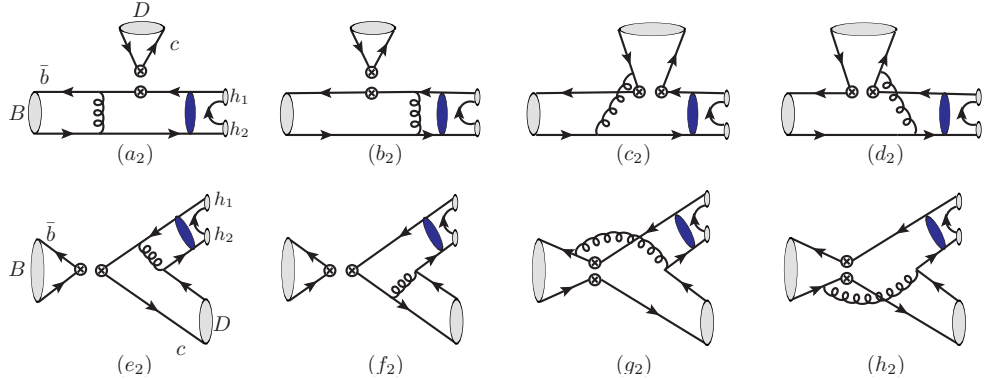}}
\vspace{-16.7cm}
\caption{ The leading-order Feynman diagrams for the quasi-two-body decays $B_{(s)} \to  D_{(s)}  K^* \to D_{(s)}  K \pi$
where $B_{(s)}$ =($B^0$, $B^+$, $B^0_s$) and $D_{(s)}$=($D^0$, $D^+$, $D_s^+$). The $h_1h_2$ denotes the $K\pi$ pair and the blue
ellipse represents the intermediate state $K^*$.}
\label{fig:fig2}
\vspace{-1cm}
\end{center}
\end{figure}

The typical Feynman diagrams at the
 leading order for the quasi-two-body decays $B_{(s)} \to  \bar{D}_{(s)}  K^* \to \bar{D}_{(s)}  K \pi$ (through $\bar{b} \to \bar{c}$
 transition) and $B_{(s)} \to  D_{(s)}  K^* \to D_{(s)}  K \pi$ (through $\bar{b} \to \bar{u}$ transition) are shown in Fig.~\ref{fig:fig1} and
 \ref{fig:fig2}, respectively. By making analytical evaluations for those Feynman diagrams in Fig.~\ref{fig:fig1} and Fig.~\ref{fig:fig2}, we
 can obtain the total decay amplitudes of these concerned decays.


For the $B_{(s)} \to \bar{D}_{(s)} K^* \to \bar{D}_{(s)} K\pi$ decays, their total decay amplitudes
can be written explicitly in the following form
\beq
{\cal A}(B^+ \to \bar{D}^0K^{*+} \to \bar{D}^0K \pi) &=& \frac{G_F} {\sqrt{2}}
V_{cb}^*V_{us}[(C_1+\frac{C_2}{3})F^{LL}_{eK^{*}}+C_2M^{LL}_{eK^{*}}+(\frac{C_1}{3}+C_2)F^{LL}_{eD}+C_1M^{LL}_{eD}] \;, \label{amp1}
\eeq

\beq
{\cal A}(B^0 \to \bar{D}^0K^{*0} \to \bar{D}^0K \pi) &=& \frac{G_F} {\sqrt{2}}
V_{cb}^*V_{us}[(C_1+\frac{C_2}{3})F^{LL}_{eK^{*}}+C_2M^{LL}_{eK^{*}}] \;, \label{amp2}
\eeq

\beq
{\cal A}(B^0 \to D^- K^{*+} \to D^-K \pi) &=& \frac{G_F} {\sqrt{2}} V_{cb}^*V_{us}[(\frac{C_1}{3}+C_2)F^{LL}_{eD}+C_1M^{LL}_{eD}] \;,
\label{amp3}
\eeq

\beq
{\cal A}(B^0 \to D_s^- K^{*+} \to D_s^- K \pi) &=& \frac{G_F} {\sqrt{2}} V_{cb}^*V_{ud}[(C_1+\frac{C_2}{3})F^{LL}_{aK^{*}}+C_2M^{LL}_{aK^{*}}]
\;, \label{amp4}
\eeq

\beq
{\cal A}(B_s^0 \to \bar{D}^0 \bar{K}^{*0} \to \bar{D}^0K \pi) &=& \frac{G_F} {\sqrt{2}}
V_{cb}^*V_{ud}[(C_1+\frac{C_2}{3})F^{LL}_{eK^{*}}+C_2M^{LL}_{eK^{*}}] \;, \label{amp5}
\eeq

\beq
{\cal A}(B_s^0 \to D_s^- K^{*+} \to D_s^-  K \pi) &=& \frac{G_F} {\sqrt{2}}
V_{cb}^*V_{us}[(\frac{C_1}{3}+C_2)F^{LL}_{eD}+C_1M^{LL}_{eD}+(C_1+\frac{C_2}{3})F^{LL}_{aK^{*}}+C_2M^{LL}_{aK^{*}}] \;, \label{amp6}
\eeq
while  the total decay amplitudes for $B_{(s)} \to D_{(s)} K^* \to D_{(s)} K\pi$ decays can be written as
\beq
{\cal A}(B^+ \to D^0 K^{*+} \to D^0K \pi) &=& \frac{G_F} {\sqrt{2}}
V_{ub}^*V_{cs}[(C_1+\frac{C_2}{3})F^{LL}_{eK^{*}}+C_2M^{LL}_{eK^{*}}+(\frac{C_1}{3}+C_2)F^{LL}_{aD}+C_1M^{LL}_{aD}] \;, \label{amp7}
\eeq

\beq
{\cal A}(B^+ \to D^+ K^{*0} \to D^+ K \pi) &=& \frac{G_F} {\sqrt{2}} V_{ub}^*V_{cs}[(\frac{C_1}{3}+C_2)F^{LL}_{aD}+C_1M^{LL}_{aD}] \;,
\label{amp8}
\eeq

\beq
{\cal A}(B^+ \to D_s^+ \bar{K}^{*0} \to D_s^+K \pi) &=& \frac{G_F} {\sqrt{2}} V_{ub}^*V_{cd}[(\frac{C_1}{3}+C_2)F^{LL}_{aD}+C_1M^{LL}_{aD}] \;,
\label{amp9}
\eeq

\beq
{\cal A}(B^0 \to D^0 K^{*0} \to D^0K \pi) &=& \frac{G_F} {\sqrt{2}} V_{ub}^*V_{cs}[(C_1+\frac{C_2}{3})F^{LL}_{eK^{*}}+C_2M^{LL}_{eK^{*}}] \;,
\label{amp10}
\eeq

\beq
{\cal A}(B^0 \to D_s^+ K^{*-} \to D_s^+K \pi) &=& \frac{G_F} {\sqrt{2}} V_{ub}^*V_{cd}[(C_1+\frac{C_2}{3})F^{LL}_{aD}+C_2M^{LL}_{aD}] \;,
\label{amp11}
\eeq

\beq
{\cal A}(B_s^0 \to D^0 \bar{K}^{*0} \to D^0 K \pi) &=& \frac{G_F} {\sqrt{2}}
V_{ub}^*V_{cd}[(C_1+\frac{C_2}{3})F^{LL}_{eK^{*}}+C_2M^{LL}_{eK^{*}}] \;, \label{amp12}
\eeq

\beq
{\cal A}(B_s^0 \to D^+ K^{*-} \to D^+K \pi) &=& \frac{G_F} {\sqrt{2}} V_{ub}^*V_{cd}[(\frac{C_1}{3}+C_2)F^{LL}_{eK^{*}}+C_1M^{LL}_{eK^{*}}] \;,
\label{amp13}
\eeq

\beq
{\cal A}(B_s^0 \to D_s^+ K^{*-} \to D_s^+K \pi) &=& \frac{G_F} {\sqrt{2}}
V_{ub}^*V_{cs}[(\frac{C_1}{3}+C_2)F^{LL}_{eK^{*}}+C_1M^{LL}_{eK^{*}}+(C_1+\frac{C_2}{3})F^{LL}_{aD}+C_2M^{LL}_{aD}] \;, \label{amp14}
\eeq
where the individual amplitude $F_{eK^{*}}^{LL},$
$M_{eK^{*}}^{LL},$ $F_{eD}^{LL},$ $M_{eD}^{LL}, F_{aK^{*}}^{LL}$ and $M_{aK^{*}}^{LL}$ are the amplitudes from
different sub-diagrams in Fig.~\ref{fig:fig1} and Fig.~\ref{fig:fig2}. Since the $P$-wave kaon-pion distribution amplitudes in
Eq.~(\ref{eq:phifunc}) have the same Lorentz structure as that of two-pion ones in Ref.~\cite{npb923-54,plb763-29},
the concerned expressions of those individual amplitudes in Ref.~\cite{npb923-54}
can be employed in this work directly by replacing the
distribution amplitudes ($\phi_0$, $\phi_s$, $\phi_t$) of the $\pi\pi$ system with the corresponding twists of the $K\pi$ ones in
Eq.~({\ref{eq:phi1}})-(\ref{eq:phi3}). The
parameter $c$ in the Eq.~(33) of~\cite{npb923-54} is adopted to be $0.4$ in this work according to the
Refs.~\cite{prd65-014007,prd80-074024}.

For the $B \to  D   K^*  \to  D K \pi$ decays, the differential decay rate can be described as
\beq
\frac{d{\cal B}}{d\omega^2}=\tau_{B}\frac{|\vec{p}_1|
|\vec{p}_3 | }{64\pi^3m^3_{B}}|{\cal A}|^2,
\label{expr-br}
\eeq
where $\tau_{B}$ is the mean lifetime of $B$ meson, the kinematic variables $|\vec{p}_1|$ and $|\vec{p}_3|$ denote the magnitudes of
the $K$ and $D$ momenta in the center-of-mass frame of the kaon-pion pair,
\beq
|\vec{p}_1|&=&\frac{1}{2} \sqrt{[(m^2_{K}-m^2_\pi)^2-2(m^2_{K}+m^2_\pi)w^2+w^4]/w^2}, \non
|\vec{p}_3|&=&\frac{1}{2} \sqrt{[(m^2_{B}-m^2_D)^2-2(m^2_{B}+m^2_D)w^2+w^4]/w^2}.
\eeq

\section{Numerical results and Discussions}\label{sec:3}
 The adopted input parameters in our numerical calculations are summarized as following (the masses, decay constants and QCD scale are in units
 of {\rm GeV})~\cite{PDG2018}:
\beq
\Lambda^{(f=4)}_{ \overline{MS} }&=&0.25,\quad m_{B^+}=5.279,\quad m_{B^0}=5.280,\quad m_{B_s}=5.367,\quad m_{D^\pm}=1.870,\nonumber\\
\quad m_{D^0/\bar{D}^0}&=&1.865,\quad m_{D_s^\pm}=1.968,
\quad m_{b}=4.8,\quad m_{c}=1.275,\quad m_{\pi^\pm}=0.140,\nonumber\\
 \quad m_{\pi^0}&=&0.135,\quad m_{K^\pm}=0.494,\quad m_{K^0/\bar{K}^0}=0.498,\quad f_{B}=0.19,\quad f_{B_s}=0.236,\nonumber\\
\quad f_{D}&=&0.2119, \quad f_{D_s}=0.249,
\quad \tau_{B^+}= 1.638\; {\rm ps},\quad \tau_{B^0}= 1.520\; {\rm ps},\quad \tau_{B_s^0}= 1.509\; {\rm ps}. \label{eq:inputs}
\eeq
For the Wolfenstein parameters $(A, \lambda,\bar{\rho},\bar{\eta})$ of the CKM mixing matrix, we use
$A=0.836\pm0.015,~\lambda=0.22453\pm0.00044$,~$\bar{\rho} = 0.122^{+0.018}_{-0.017},~\bar{\eta}= 0.355^{+0.012}_{-0.011}$~\cite{PDG2018}.

By using the decay amplitudes as given in Eq.~(\ref{amp1}-\ref{amp14}) and
the differential branching ratio in Eq.~(\ref{expr-br}),
integrating over the full $K\pi$ invariant mass region  $(m_K+m_{\pi})\leq \omega \leq (m_B-m_D)$
 for the resonant components,
we obtain the branching ratios for the quasi-two-body decays
$B_{(s)} \to  \bar{D}_{(s)}   K^* \to  \bar{D}_{(s)}  K \pi$ and
$B_{(s)} \to  D_{(s)}   K^* \to  D_{(s)}  K \pi$, and list
the numerical results in Table~\ref{tab1} and Table~\ref{tab2}.
The first error of these PQCD predictions comes from the $B(B_{s})$ meson shape parameter uncertainty
$\omega_B = 0.40 \pm 0.04$~($\omega_{B_s}=0.50 \pm 0.05$) {\rm GeV}, the following two errors are from
the Gegenbauer coefficients in the kaon-pion distribution
amplitudes: $a_{1K^*}^{||}=0.05\pm0.02$, $a_{2K^*}^{||}=0.15\pm0.05$ and the last one is induced by
$C_D=0.5 \pm 0.1$~($C_{D_s}= 0.4\pm 0.1$) for $D(D_s)$ meson wave function.
One can see that the dominant theoretical error comes from the
uncertainty of $\omega_{B_{(s)}}$: about 10\% for the decays $B^+ \to D^+ K^{*0} \to D^+K \pi$
and $B^+ \to D_s^+ \bar{K}^{*0} \to D_s^+K \pi$, and about 20\% to 40\% for other remaining decays. It can be roughly understood from the
analytic formulas of Eq.~(\ref{amp8}) and (\ref{amp9}). Since $C_2$ is much
 more than $C_1$, the main contribution for those two pure annihilation type decays
$B^+ \to D^+ K^{*0} \to D^+K \pi$ and $B^+ \to D_s^+ \bar{K}^{*0} \to D_s^+K \pi$ comes from the term which is proportional to
$F^{LL}_{aD}$, the amplitude of the factorizable annihilation diagram. While the amplitude $F^{LL}_{aD}$ does not contain any term
about $\omega_{B_{(s)}}$, the corresponding
theoretical error due to $\omega_{B_{(s)}}$ is reduced for those two decays naturally.
 One can also see that the error stemming from $a_{1K^*}^{||}, a_{2K^*}^{||}$
and $C_{D_{(s)}} $ is less than 15\% respectively.
The errors come from the uncertainties
of the parameters, for instance, the Wolfenstein parameters, the pole mass $m_{K^*}$ and width $\Gamma_{K^*}$, are very
small and have been neglected.
Although the three-body $B$ meson decay offers an ideal ground to study the distribution of $CP$ asymmetry,
there are no direct $CP$ violations for these decays in this work because only the tree diagrams contribute to the considered
decay processes.

In this work,
the Gegenbauer moments for $K^*$ at the scale of $1$ GeV are chosen, similar to the definitions for the ordinary
distribution amplitudes used in the PQCD approach. In principle, the Gegenbauer moments should depend
on the factorization scale. As a test of the effect of the scale evolution for the Gegenbauer moments in the kaon-pion distribution
amplitudes, we  calculate the branching ratios of  the decays $B^+ \to \bar{D}^0 K^{*+} \to \bar{D}^0 K \pi$
 and $B^0 \to D^- K^{*+} \to D^-K \pi$
 by considering the $a_{1K^*}^{||}, a_{2K^*}^{||}$ with the evolution from $1$ {\rm GeV} to the hard scale $\sqrt{\Lambda m_B}$.
 The new results for $B^+ \to \bar{D}^0 K^{*+} \to \bar{D}^0 K \pi$
 and $B^0 \to D^- K^{*+} \to D^-K \pi$ are $5.28 \times 10^{-4}$ and  $3.51 \times 10^{-4}$, respectively,
 and the variations are found to be less than $4\%$ for the corresponding values in Table~\ref{tab1}. Which mean that
we can neglect the scale evolution for the  kaon-pion system  Gegenbauer moments when considering that the current data for the
three-body $B$ decays still have larger uncertainties.

\begin{table}[htb]
\begin{center}
\caption{The PQCD predictions for the branching ratios of  the
$B_{(s)} \to  \bar{D}_{(s)}   K^* \to  \bar{D}_{(s)}  K \pi$ decays in the quasi-two-body framework together with experimental data.}
\label{tab1}
\begin{tabular}{ l  c c  l } \hline\hline
{\rm ~~~~~~~~~~~~~~~~~~~~Mode} & {\rm Unit}& {\rm ${\cal B}_{th}$} &~~~~~~~~~~~~~~~~~~~~~~~~~~ {\rm ${\cal B}_{exp}$}  \\ \hline\hline
$B^+ \to \bar{D}^0 K^{*+} \to \bar{D}^0 K \pi~~~~$&$(10^{-4})$&$~~~~5.38^{+1.75+0.04+0.12+0.41}_{-1.55-0.45-0.39-0.56}$&
~~~~BABAR\cite{prd73-111104}: $5.29 \pm0.30 \pm0.34 $    \\ \hline
$B^0 \to \bar{D}^0 K^{*0} \to \bar{D}^0K \pi~~~~$ &$(10^{-5})$&
$~~~~1.93^{+0.60+0.03+0.22+0.01}_{-0.33-0.07-0.18-0.05}$&~~~~Belle\cite{prl90-141802}: $4.8^{+1.1}_{-1.0}\pm0.5$ \\
$ $ & $ $&$ $&~~~~BABAR\cite{prl96-011803}: $5.7\pm0.9\pm0.6$ \\
$ $ & $ $&$ $&~~~~BABAR\cite{prd74-031101}: $4.0\pm0.7\pm0.3$ \\
$ $ &$ $ &$ $&~~~~LHCb\cite{prd92-012012}: $5.13\pm0.20\pm0.15\pm0.24\pm0.60$ \\ \hline
$B^0 \to D^- K^{*+} \to D^-K \pi~~~~$ &$(10^{-4})$&
$~~~~3.63^{+1.61+0.04+0.02+0.35}_{-0.91-0.05-0.11-0.36}$&~~~~BABAR\cite{prl95-171802}: $4.6 \pm 0.6 \pm 0.5$ \\  \hline
$ B^0 \to D_s^- K^{*+} \to D_s^-K \pi~~~~$ &$(10^{-4})$& $~~~~1.05^{+0.18+0.03+0.07+0.15}_{-0.06-0.03-0.01-0.07}$
&~~~~BABAR\cite{prd78-032005}: $0.35^{+0.10}_{-0.09}\pm 0.04$\\ \hline
$ B_s^0 \to \bar{D}^0 \bar{K}^{*0} \to \bar{D}^0K \pi~~~~$ &$(10^{-4})$&$~~~~2.33^{+0.94+0.31+0.60+0.25}_{-0.66-0.13-0.11-0.07}$
&~~~~LHCb\cite{plb706-32}: $4.72\pm1.07\pm0.48\pm0.37\pm0.74$ \\
$ $ &$$&$$
&~~~~LHCb\cite{plb727-403}: $3.3\pm0.3\pm0.1\pm0.3\pm0.5$ \\
$ $ &$$&$$
&~~~~LHCb\cite{prd90-072003}: $4.29\pm0.09\pm0.11\pm0.14\pm0.63$ \\ \hline
$ B_s^0 \to D_s^- K^{*+} \to D_s^-K \pi~~~~ $ &$(10^{-4})$&$~~~~2.06^{+0.74+0.02+0.01+0.17}_{-0.53-0.02-0.01-0.16}$ &$ $  \\
 \hline\hline
\end{tabular} \end{center}
\end{table}

\begin{table}[htb]
\begin{center}
\caption{The PQCD predictions for the branching ratios of  the CKM suppressed $B_{(s)} \to  D_{(s)}   K^* \to  D_{(s)} K \pi$ decays
in the quasi-two-body framework together with experimental data.}
\label{tab2}
\begin{tabular}{ l  c c  l } \hline\hline
{\rm ~~~~~~~~~~~~~~~~~~~~Mode} & {\rm Unit}& {\rm ${\cal B}_{th}$} &~~~~~~~~~~~~~~~~~~~~~~~~~~ {\rm ${\cal B}_{exp}$}  \\ \hline\hline
$B^+ \to D^0 K^{*+} \to D^0  K \pi~~~~$&$(10^{-6})$&$~~~~4.49^{+1.72+0.32+0.58+0.01}_{-0.82-0.15-0.59-0.05} $& $ $ \\\hline
$B^+ \to D^+ K^{*0} \to D^+K \pi~~~~$&$(10^{-7})$ &$~~~~6.27^{+0.58+0.14+0.04+0.39}_{-0.79-0.10-0.04-0.35} $& ~~~~BABAR\cite{prd82-092006}: $<
 30$ at 90\% C.L.\\
$ $ & $ $&$ $&~~~~LHCb\cite{jhep1302-043}: $<  18$ at 90\% C.L. \\
$ $ & $ $&$ $&~~~~LHCb\cite{prd93-051101}: $<  4.9(6.1)$ at 90(95)\% C.L. \\\hline
$B^+ \to D_s^+ \bar{K}^{*0} \to D_s^+K \pi~~~~$&$(10^{-8})$ &
$~~~~3.80^{+0.23+0.13+0.06+0.25}_{-0.43-0.13-0.28-0.32} $& ~~~~LHCb\cite{jhep1302-043}: $<  440$ at 90\% C.L.\\\hline
$ B^0 \to D^0 K^{*0} \to D^0K \pi~~~~$ &$(10^{-6})$& $~~~~3.25^{+1.06+0.18+0.36+0.08}_{-0.68-0.38-0.13-0.04} $ &~~~~Belle\cite{prl90-141802}:
$<18$ at 90\% C.L.\\
$ $ & $ $&$ $&~~~~BABAR\cite{prd74-031101}: $<11$ at 90\% C.L. \\\hline
$ B^0 \to D_s^+ K^{*-} \to D_s^+K \pi~~~~$&$(10^{-8})$ &$~~~~3.08^{+0.42+0.31+0.27+0.15}_{-0.43-0.22-0.46-0.21} $&$ $ \\\hline
$ B_s^0 \to D^0 \bar{K}^{*0} \to D^0 K \pi~~~~ $&$(10^{-7})$ &$~~~~1.15^{+0.43+0.01+0.19+0.07}_{-0.36-0.05-0.22-0.12} $&$ $  \\\hline
$ B_s^0 \to D^+ K^{*-} \to D^+K \pi~~~~ $&$(10^{-6})$ &$~~~~2.00^{+0.75+0.04+0.11+0.03}_{-0.51-0.04-0.11-0.02} $&$ $  \\\hline
$ B_s^0 \to D_s^+ K^{*-} \to D_s^+K \pi~~~~ $&$(10^{-5})$ &$~~~~4.79^{+1.82+0.07+0.12+0.01}_{-1.26-0.06-0.12-0.07} $&$ $  \\
 \hline\hline
\end{tabular} \end{center}
\end{table}

From the calculations and the numerical results as listed in Table~\ref{tab1} and Table~\ref{tab2}, one can find the following points:
\begin{itemize}
\item[(1)]
By assuming the ${\cal B}(K^*\to K\pi) \approx 100$ \%~\cite{PDG2018} and accepting a simple relation between the  branching ratio of the same kind of decay evaluated in the quasi-two-body and the two-body framework, it is easy to have
\beq
{\cal B}(B \to  D  K^* \to D  K \pi)&=& {\cal B}(B \to  D  K^*) \cdot
{\cal B}(K^* \to K \pi) \approx {\cal B}(B \to  D  K^*).
\eeq
\item
This relation can be examined roughly from the predictions in this work and the results
calculated in the PQCD framework for two-body decays~\cite{prd78-014018,jpg37-015002}.
For examples, ${\cal B}(B^+ \to \bar{D}^0 K^{*+} \to \bar{D}^0 K \pi)=5.38\times 10^{-4}$
and ${\cal B}(B_s^0 \to D_s^- K^{*+} \to D_s^-K \pi)=2.06\times 10^{-4}$ in Table~\ref{tab1}
consistent with the
${\cal B}(B^+ \to \bar{D}^0 K^{*+})=6.37\times 10^{-4}$ and ${\cal B}(B_s^0 \to D_s^- K^{*+} )=2.81\times 10^{-4}$
in the Ref.~\cite{prd78-014018}. While because of the updated parameters, there are also relatively large differences between
the results for some decays,
 especially for those channels with the branching ratios less than $10^{-5}$. The same situation exists in the relevant decays with
 the resonance $\rho$ since ${\cal B}(\rho \to \pi\pi) \approx 100$ \%. With the same input parameters, the PQCD predictions for the branching ratios of all considered decays obtained in both the quasi-two-body and
the two-body decay frameworks were found to agree very well with each other~\cite{npb923-54}.

\item
One can extract the PQCD predictions for the decay rates of the related quasi-two-body decays from the results in
Table~\ref{tab1} and Table~\ref{tab2}.
Take the quasi-two-body decay $B^+ \to \bar{D}^0 K^{*+} \to \bar{D}^0 K^+ \pi^0$ for example, the relation between ${\cal
B}(B^+ \to \bar{D}^0 K^{*+} \to \bar{D}^0 K^+ \pi^0)$ and ${\cal B}(B^+ \to \bar{D}^0 K^{*+})$ can be described as
\beq
{\cal B}(B^+ \to \bar{D}^0 K^{*+} \to \bar{D}^0 K^+ \pi^0)&=& {\cal B}(B^+ \to \bar{D}^0 K^{*+}) \cdot {\cal B}(K^{*+} \to K^+
\pi^0),
\eeq
where the isospin relation ${\cal B}(K^{*+} \to K^+ \pi^0)=\frac{1}{3}$.
Combining with the central value of ${\cal B}(B^+ \to \bar{D}^0 K^{*+} \to \bar{D}^0 K \pi)=5.38\times 10^{-4}$ in Table~\ref{tab1},
one can obtain the PQCD prediction for ${\cal B}(B^+ \to \bar{D}^0 K^{*+} \to \bar{D}^0
K^+\pi^0)=1.79\times 10^{-4}$ easily.

\item[(2)]
Compare our numerical results with the experimental data in Table~\ref{tab1}, one can see that:
\item
The PQCD prediction for the branching
 ratio of the decay $B^+ \to \bar{D}^0 K^{*+} \to \bar{D}^0 K \pi$  agrees well with the value
$(5.29 \pm0.30 \pm0.34) \times 10^{-4}$~\cite{prd73-111104} measured by
the ${\it BABAR}$ detector at the PEP-II $B$ Factory.  In Ref.~\cite{prl95-171802}, the branching
 ratio of $B^0 \to D^- K^{*+}$ decay was measured to be  $(4.6 \pm 0.6 \pm 0.5)\times
 10^{-4}$ and the $K^*$ resonant fraction $\frac{{\cal B}(B^0 \to D^-K^{*+}){\cal
B}(K^{*+} \to K^0 \pi^{+})}{{\cal B}(B^0 \to D^- K^0 \pi^{+})}=0.63 \pm0.08\pm0.04$ has been also obtained.
${\cal B}(B^0 \to D^-K^{*+}\to D^-K\pi)=(3.63^{+1.61+0.04+0.02+0.35}_{-0.91-0.05-0.11-0.36})\times
 10^{-4}$ predicted in this work agrees with the result within the measurement uncertainties. We also have the PQCD prediction
 $\frac{{\cal B}(B^0 \to D^-K^{*+}){\cal
B}(K^{*+} \to K^0 \pi^{+})}{{\cal B}(B^0 \to D^- K^0 \pi^{+})}\approx 0.5$
by taking ${\cal B}(B^0 \to D^- K^0 \pi^{+})=(4.9 \pm0.7 \pm0.5) \times 10^{-4}$ from Ref.~\cite{prl95-171802}.

\item
For the decay $B^0 \to \bar{D}^0 K^{*0} \to \bar{D}^0K \pi$, the central value of the branching fraction predicted by PQCD is
less than $50$\% of the available experimental measurements but agrees with the PQCD prediction in the ordinary two-body
framework in~\cite{prd78-014018} within errors. The prediction ${\cal B}(B_s^0 \to \bar{D}^0 \bar{K}^{*0} \to \bar{D}^0K
\pi)=(2.33^{+0.94+0.31+0.60+0.25}_{-0.66-0.13-0.11-0.07})\times 10^{-4}$ in this work is more close to $(3.3\pm0.3\pm0.1\pm0.3\pm0.5)\times
10^{-4}$ in Ref.~\cite{plb727-403} than other two results from LHCb~\cite{plb706-32,prd90-072003}. Furthermore, in
Ref.~\cite{plb706-32}, the measurement of the ratio of branching fractions for the decays $B_s^0 \to \bar{D}^0 \bar{K}^{*0}$ and $B^0
\to \bar{D}^0 \rho^0$ was found to be
\beq
\frac{{\cal B}(B_s^0 \to \bar{D}^0 \bar{K}^{*0})}{{\cal B}(B^0 \to \bar{D}^0 \rho^0)}=1.48\pm0.34\pm0.15\pm0.12,
\eeq
while a similar ratio between ${\cal B}(B_s^0 \to \bar{D}^0 \bar{K}^{*0})$ and ${\cal B}(B^0 \to \bar{D}^0 K^{*0})$ was measured in
Ref.~\cite{plb727-403}
\beq
\frac{{\cal B}(B_s^0 \to \bar{D}^0 \bar{K}^{*0})}{{\cal B}(B^0 \to \bar{D}^0 K^{*0})}=7.8\pm0.7\pm0.3\pm0.6.
\eeq
Utilize the PQCD predictions in Table~\ref{tab1} and ${\cal B}(B^0 \to \bar{D}^0 \rho^0)=1.39\times 10^{-4}$ taken from our previous
work in Ref.~\cite{npb923-54}, we estimate the ratio of branching fractions $\frac{{\cal B}(B_s^0 \to \bar{D}^0 \bar{K}^{*0})}{{\cal B}(B^0
\to\bar{D}^0 \rho^0)}=1.68$,
 in agreement with the data, while our prediction $\frac{{\cal B}(B_s^0 \to \bar{D}^0 \bar{K}^{*0})}{{\cal B}(B^0
\to \bar{D}^0 K^{*0})}=12.07$ is a bit larger than the value announced by LHCb~\cite{plb727-403} after taking errors
into consideration. If we accept ${\cal B}(B_s^0 \to \bar{D}^0 \bar{K}^{*0})=4.29 \times 10^{-4}$~\cite{prd90-072003} ($4.72 \times
10^{-4}$~\cite{plb706-32}) and ${\cal B}(B^0 \to \bar{D}^0 K^{*0})= 4.0 \times 10^{-5}$ in Ref.~\cite{prd74-031101}, our result for the
ratio $\frac{{\cal B}(B_s^0 \to \bar{D}^0 \bar{K}^{*0})}{{\cal B}(B^0 \to \bar{D}^0 K^{*0})}$ is still consistent with the data.

\item
The branching fraction for the pure annihilation decay $B^0 \to D_s^- K^{*+}$: ${\cal B}=(0.35^{+0.10}_{-0.09}\pm 0.04)\times 10^{-4}$
selected from ${\it BABAR}$~\cite{prd78-032005} is smaller than the values $1.05\times
10^{-4}$ obtained from this work and  $1.82\times 10^{-4}$ acquired from the PQCD framework for two-body
decays~\cite{prd78-014018}. More theoretical studies and experimental measurements are needed to improve the estimation for the pure annihilation decays.

\item[(3)]
For the  branching ratios of eight CKM suppressed $B_{(s)} \to  D_{(s)}   K^* \to  D_{(s)} K \pi$ decays as listed in Table~\ref{tab2},
the PQCD predictions are in the order of $10^{-8}$ to $10^{-5}$ and we have some comments as follows:
\item
 Since no enough significant signals have been observed,
there were hardly any specific
data for the branching fractions of those decays but the upper limits. As shown in Table~\ref{tab2},
the experimental collaborations have determined the upper limits for three of
the considered decays at $90(95)$\% confidence level and it is easy to see that all the PQCD predictions for the branching ratios are
consistent with the corresponding experimental ones.

\item
In Ref.~\cite{prd93-051101}, the ratio of branching fractions was measured to be
\beq
\frac{{\cal B}(B^+ \to D^+ K^{*0} \to D^+ K^+ \pi^-)}{{\cal B}(B^+ \to  D^- K^+ \pi^+)}< 0.0044(0.0055),
\eeq
at $90(95)$\% confidence level. By adopting the measured value ${\cal B}(B^+ \to  D^- K^+ \pi^+)=(7.31\pm0.19\pm0.22\pm0.39)\times
10^{-5}$~\cite{prd91-092002}, the isospin relation ${\cal B}(K^{*0} \to K^+ \pi^-)=\frac{2}{3}$, and our result ${\cal B}(B^+ \to D^+ K^{*0}
\to DK \pi)=(6.27^{+0.58+0.14+0.04+0.39}_{-0.79-0.10-0.04-0.35})\times 10^{-7}$, we estimate the ratio  $\frac{{\cal B}(B^+ \to D^+ K^{*0}
\to D^+ K^+ \pi^-)}{{\cal B}(B^+ \to  D^- K^+ \pi^+)}\approx 0.0057$. It is to be tested further since
 the upper limit for the branching fraction
of $B^+ \to D^+ K^{*0} \to DK \pi$ given by Ref.~\cite{prd93-051101} is much less than that in the
Ref.~\cite{prd82-092006} and Ref.~\cite{jhep1302-043}.

\item
We suggest more studies for those CKM suppressed $B_{(s)} \to  D_{(s)}   K^*
\to  D_{(s)} K \pi$ decays in which the decay mode $ B_s^0 \to D_s^+ K^{*-} \to D_s^+K \pi$ has a large branching ratio,
and could be measured in the LHCb and Belle-II experiments.

\begin{figure}[tb]
\vspace{-0.5cm}
\centerline{\epsfxsize=7.5cm \epsffile{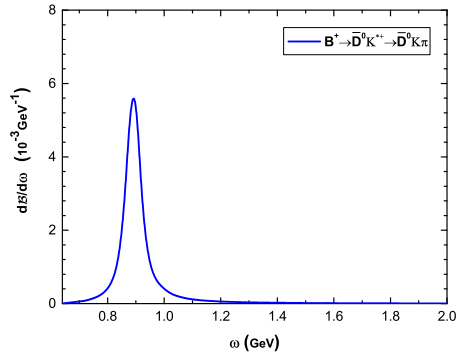} }
\vspace{-0.5cm}
\caption{The PQCD prediction for the $\omega$-dependence of the differential branching ratio for the decay mode
$B^+ \to \bar{D}^0 K^{*+} \to \bar{D}^0 K \pi$. }\label{fig3}
\end{figure}
\item[(4)]
Different from the fixed kinematics of the two-body $B$ meson decays, the decay amplitudes of the quasi-two-body $B$ meson
decays show a strong dependence on the  $K\pi$ invariant mass $\omega$. In Fig.~\ref{fig3}, we plot the differential decay branching
ratio of the decay mode
$B^+ \to \bar{D}^0 K^{*+} \to \bar{D}^0 K \pi$ versus the invariant mass $\omega$
in the range of $[m_K+m_{\pi}, 2 {\rm GeV}]$. The main portion
 of the branching ratio lies obviously in the region around the pole mass of the resonant state which presented as a narrow peak in the
 plot, the contributions from the energy region $\omega>1.2$ {\rm GeV} can be omitted safely.
\end{itemize}

\section{Summary}\label{sec:4}
Motivated by the abundant experimental data, we studied the contributions of the $P$-wave resonant
states $K^*$ to the decays $B_{(s)} \to   \bar{D}_{(s)} K \pi$
and the CKM suppressed decays $B_{(s)} \to   D_{(s)}K \pi$ by employing the PQCD factorization approach.
The final-state interactions between the $K\pi$ pair are factorized into
the two-meson distribution amplitudes in which the resonant line shape for the resonance $K^*$ in the time-like form factor
$F_{K\pi}$ is described by the RBW function.

By the numerical evaluations and the phenomenological analyses, we found the following points:
\begin{itemize}
\item[(1)]
By adopting the ${\cal B}(K^*\to K\pi) \approx 100$ \%, one can obtain $
{\cal B}(B \to  D  K^* \to D  K \pi) \approx {\cal B}(B \to  D  K^*)$ easily, and it provides us
a new way to study those two-body $B$ meson decays in the framework for the quasi-two-body cases.

\item[(2)]
The PQCD predictions for the branching fractions of the decay processes $B_{(s)} \to   \bar{D}_{(s)} K^* \to \bar{D}_{(s)}K \pi$
do  agree with the experimental data,  except for the cases of the color-suppressed decay
$B^0 \to \bar{D}^0 K^{*0} \to \bar{D}^0K \pi$ and the pure annihilation decay $B^0 \to D_s^- K^{*+} \to D_s^- K\pi$.
More precise data from the LHCb and the Belle-II experiments
can help us to test our predictions and  to improve the theoretical framework itself.

\item[(3)]
For the CKM suppressed $B_{(s)} \to  D_{(s)}   K^* \to  D_{(s)} K \pi$ decays,  all the PQCD
predictions for the branching ratios are consistent with currently  available experimental measurements.
 Experimentally, the upper limit for the branching fraction
of $B^+ \to D^+ K^{*0} \to DK \pi$ given by Ref.~\cite{prd93-051101} is much less than that in the
Ref.~\cite{prd82-092006} and Ref.~\cite{jhep1302-043}, and it is to be verified further.

 \item[(4)]
Unlike the fixed kinematics of the two-body $B$ meson decays, the decay amplitudes of the quasi-two-body $B$ meson
decays have a strong dependence on the  $K\pi$ invariant mass $\omega$ and the main portion
 lies in the region around the pole mass of the resonant state.
\end{itemize}

\begin{acknowledgments}

Many thanks to Hsiang-nan Li and Rui Zhou for valuable discussions.
This work was supported by the National Natural Science Foundation of China under the Grant No.~11775117 and  11547038.
 Ai-Jun Ma was also supported by the Scientific Research Foundation of Nanjing Institute of Technology under Grant No.~YKJ201854.

\end{acknowledgments}


\end{document}